 \documentclass{emulateapj}

\newcommand{\myemail}{Duncan.Galloway@sci.monash.edu.au}
\newcommand{\src}{EXO~0748$-$676}
\newcommand{\xte}{{\it RXTE}}

\newcommand{\epcs}{{\rm erg\,cm^{-2}\,s^{-1}}}

\slugcomment{Accepted for ApJ Letters}

\shorttitle{A 552~Hz burst oscillation in EXO~0748$-$676}
\shortauthors{Galloway et al.}

\begin{document}

\title{Discovery of a 552~Hz burst oscillation in the low-mass X-ray
binary EXO~0748$-$676}

\author{Duncan K. Galloway\altaffilmark{1}}
\affil{Center for Stellar and Planetary Astrophysics,
  Monash University, VIC 3800, Australia}

\author{Jinrong Lin\altaffilmark{2},
Deepto Chakrabarty\altaffilmark{2}} 
\affil{Kavli Institute for Astrophysics and Space Research, Massachusetts
Institute of Technology, Cambridge MA 02139}

\and

\author{Jacob M. Hartman\altaffilmark{3}} 
\affil{Space Science Division, Code 7655, Naval
  Research Laboratory, Washington DC 20375}

\email{\myemail}

\altaffiltext{1}{Monash Fellow}
\altaffiltext{2}{also Department of Physics, Massachusetts Institute of
Technology, Cambridge, MA 02139}
\altaffiltext{3}{also National Research Council Research Associate.
Current address: National
Radio Astronomy Observatory, 1003 Lopezville Road, Socorro, NM
87801}

\begin{abstract}
We report the detection of pulsations at 552~Hz in the rising phase of two
type-I (thermonuclear) X-ray bursts observed from 
the accreting neutron star
EXO~0748$-$676
in 2007 January and December,
by the {\it Rossi X-ray Timing Explorer}. 
The fractional amplitude 
was 
15\% (rms). 
The dynamic power density spectrum for each burst revealed an increase in
frequency of $\approx1$--2~Hz 
while the oscillation was present.
The frequency drift,
the high significance of the detections and the almost identical signal
frequencies measured in two bursts separated by 11 months, confirms this
signal as a burst oscillation similar to those found in 13 other sources
to date. 
We thus conclude that the spin frequency in \src\ is within a few Hz of
552~Hz, rather than 45~Hz as was suggested from an earlier signal
detection by \cite{villarreal04}.
Consequently, Doppler broadening must significantly affect spectral
features arising from the neutron star surface, so that the narrow
absorption features previously reported from an {\it XMM-Newton}\/
spectrum could not have arisen there.
The origin of both the previously reported 45~Hz oscillation and the X-ray
absorption lines is now uncertain.
\end{abstract}


\section{Introduction}

Neutron stars in low-mass X-ray binaries (LMXBs) provide observational
evidence for their rapid spins reluctantly, and via increasingly diverse
phenomena. Highly coherent burst oscillations, occuring only around the
peak of thermonuclear (type-I) bursts, were the first such phenomenon to
be discovered, and since have been detected in $\approx14$ sources
\cite[e.g.][]{watts08}.
Continuous pulsations in the persistent emission occur even more
infrequently, and occur at just above the burst oscillation frequency
in those sources which exhibit both \cite[]{chak03a}. This result supports
the hypothesis that the burst oscillation frequency traces the
neutron star spin. 
Most recently, intermittent persistent pulsations have
been detected in several sources, including one previously known burst
oscillation source \cite[e.g.][]{gal07a,altamirano07,casella07}.

It is uncertain why some sources show pulsations or burst oscillations and
others do not. 
Additionally, sources which
exhibit burst oscillations do not do so in every burst. In fact, the presence of
oscillations can be as rare as 1 burst in 14 \cite[for 4U~1916$-$053;
see][]{bcatalog}. The mechanism by which the burst oscillations are
produced is yet another uncertainty. Although oscillations 
are observed at high (fractional) amplitudes early in some bursts,
consistent with a spreading ``hot spot'' model \cite[e.g.][]{stroh97b}, the
oscillations in the burst tail, when the burning must have spread to the
entire stellar surface, are harder to explain. The oscillations may
instead (or also) arise from anisotropies in the
surface brightness 
originating from hydrodynamic instabilities \cite[]{slu02} or modes excited
in the neutron star ocean (e.g. \citealt{cb00}; see also
\citealt{heyl04,pb05}).

The low-mass X-ray binary \src\ is particularly well-studied.
This transient 
was discovered during
{\it EXOSAT}\/ observations in 1985 \cite[]{parmar86}, which also revealed
the first thermonuclear bursts from the source as well as X-ray dipping
activity. 
Synchronous X-ray and optical eclipses \cite[]{crampton86} are observed
once every 3.82~hr orbit.
A variety of low- and high-frequency variability has been characterised
\cite[e.g.][]{homan99,homan00}, notably including a 695~Hz quasi-periodic
oscillation.
More recently, absorption features in the summed X-ray spectra of bursts observed
by {\it XMM-Newton}\/ were
identified as redshifted lines from near the neutron star surface
\cite[$z=0.35$;][]{cott02}. The narrowness of these features requires that
the neutron star is rotating slowly, as rotation speeds $\ga100$~Hz will
broaden the line profiles to the point where they 
are undetectable
\cite[]{ozel03,chang06,bml06}.
Subsequently, \cite{villarreal04} detected a 45~Hz peak in the
summed power spectrum of 38 thermonuclear bursts, which they interpreted
as a spin frequency sufficiently
slow to give negligible broadening.
However, subsequent followup studies have failed to
confirm the spectral line detection \cite[e.g.][]{cott08}. 

Here we present analysis of recently observed bursts from \src\ which
suggest that the neutron star spin is not
45~Hz, but 552~Hz.

\section{Observations}

We analysed observations of \src\ made 
with the Proportional Counter Array \cite[PCA;][]{xte96} 
onboard the {\it Rossi X-ray Timing
Explorer}\/ ({\it RXTE}).
The PCA consists of five
identical, co-aligned proportional counter units (PCUs), sensitive to
photons in the energy range 2--60~keV. 
A passive collimator restricts incoming photons to a circular field-of-view 
with radius
$\approx1^\circ$.  Photon counts from the PCA are processed
independently by up to 6 Event Analyzers (EAs) in a variety of
configurations,
permitting time resolution down to $1\mu s$ and up to 256 spectral channels.

In an earlier paper we presented a detailed study of the properties of all
thermonuclear bursts in public \xte\/ data through 2007 June \cite[][
hereafter G08]{bcatalog}.  Additional observations are continually being
made public, and we are periodically analysing these newly available
data\footnote{Available from the High-Energy Astrophysics Science Archive
Research Center (HEASARC),
at {\url http://heasarc.gsfc.nasa.gov}.}
to extend the G08 sample.
\src\ was observed intensively throughout 2006--7 to
monitor low- and high-frequency quasi-periodic oscillations. We
analysed 223 additional observations to the G08 sample, and
detected 67 additional thermonuclear bursts.  Our burst detection and
analysis procedures are 
identical to those of G08.

One component of the analyses is
a search for burst oscillations from sources that have not
previously exhibited them.
For each burst, 
we
extract a lightcurve over the full PCA energy range and binned on $122\mu
s$.
We then
search for oscillations 
while the burst flux is
at least 10\% of the peak value for that burst. 
For \src, the search duration in all the bursts was
typically in the range 12--90~s, or 40~s in the mean.
The search is carried out
by computing Fast Fourier Transforms (FFTs) of overlapping windows of data
of length 1, 2, and 4~s, stepped by 0.25~s. For each power spectrum we
search for excess power 
at frequencies $>10$~Hz, below which red
noise from the burst rise and decay dominates.

We set a detection
threshold corresponding to $3\sigma$ significance taking into account the
number of trials for each window. However, we consider a single detection
at this confidence level,
or detections in windows that overlap in time, insufficient to confirm 
a burst oscillation. Robust detections require exceeding our
threshold in power spectra from multiple independent time windows, either in the
same burst or other bursts from the same source, and at frequencies within
a few Hz.

\section{Results}
\label{sec1}

\begin{figure}
 \plotone{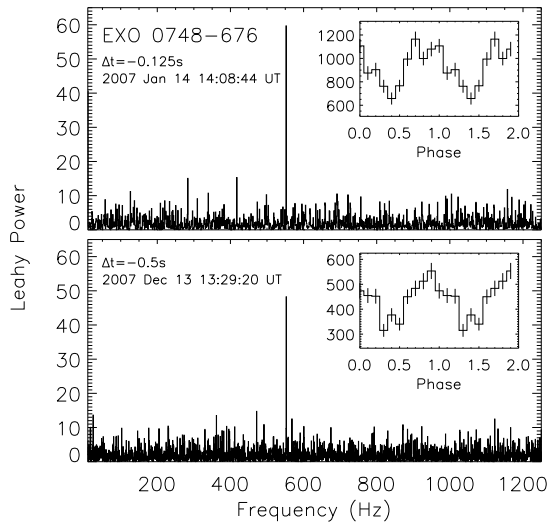}
 \figcaption[]{Significant power excess at 552~Hz in two bursts from
\src. We show the power-density spectra covering a 1-s window beginning
0.125~s before the start of the 2007 January 14 burst ({\it top
panel}), and from a 2-s window of data beginning 0.5~s before the start of
the 2007 December 13 burst ({\it bottom panel}). 
The inset
in each panel shows the folded pulse profile (units of
count~s$^{-1}$~PCU$^{-1}$) within each interval; two full cycles are
plotted for clarity. 
The relative phase of the two profiles is arbitrary.
 \label{ffts} }
\end{figure}

We measured power far exceeding the noise level in power spectra of two
bursts from \src, on 2007 January 14 14:08:44 UT (MJD~54114.58941)
and 2007 December 13 13:29:20 UT
(MJD~54447.56204)\footnote{
Observation IDs 92019-01-13-00 \&
92019-01-28-02}.
In the first burst, a
power spectrum of a 1-s window of data beginning 0.125~s before the burst start (here
defined as the time at
which the burst flux first exceeded 25\% of the peak flux, following G08)
exhibited a peak Leahy power of 59.68 at a frequency of 552~Hz (Fig.
\ref{ffts}).  The second burst, occurring almost a year later, in a 2~s
window of data beginning 0.5~s before the burst start reached a Leahy
power of 48.26 at a frequency of 552.5~Hz.

While both these detections are highly significant (with single trial
probabilities of $10^{-13}$ and $10^{-11}$) we did not
detect the signal in multiple independent (non-overlapping) time windows
in either burst.
However, the correspondence of the detection frequencies in two
separate bursts virtually guarantees the signal is real. We estimated the
likelihood of two such peaks separated by less than 1~Hz arising by
chance, as the product of the 
probabilities for each
detection, multiplied by the probability related to their separation.
Assuming a uniform distribution for noise peaks over the full frequency range of
10--4000~Hz, the probability for two such nearby detections is roughly
$2.5\times10^{-4}$. Taking into account all trials in the blind search
covering all 157 bursts from \src\ in the extended G08 catalog
the estimated 
null hypothesis probability is $10^{-10}$, equivalent to $6.3\sigma$.
This estimate includes no
correction for the lack of independence between the searches in
overlapping time windows; such correction would only increase the
significance.
We also measured the distribution of noise powers in the absence of a
signal for simulated lightcurves matching the observed 0.125-s lightcurve
within each time window in which the oscillation was detected. We
confirmed that the noise powers are distributed as $\chi^2$ with two
degrees of freedom, which supports our estimated significance.
Thus, we conclude that the signal is a genuine burst oscillation.

We folded the lightcurve in each time window in which the 552~Hz signal
was detected initially, to
measure the amplitude and harmonic content of the profile. Since the power
spectra
in each case have rather limited frequency resolution, we folded the data
on a grid of frequencies and chose the frequency value which maximised 
the variance in the folded profile.
The corresponding frequency value for the January
14 burst was 552.02~Hz, while for the December 13 burst was 552.45~Hz.
We tested for the presence of harmonic content 
from the power spectrum of the folded pulse profile, following
\cite{muno02c}. The maximum Leahy power for the 2007 January burst at $2\times$
the oscillation frequency 
was 
6.47, 
which was below our detection threshold. The power at 3 or more times the 
oscillation frequency was comparable or smaller, and even smaller powers
were measured for the December burst. Thus, we
found no evidence for significant power at the 2nd or higher harmonics
during either burst.
We fitted each profile with a sinusoidal model, giving fractional rms
amplitudes of 
$15.2\pm1.7$\% and $15\pm2$\% for the January 14 and December 13 bursts,
respectively.
The uncertainties were estimated by fixing the amplitude on
a grid of values, re-fitting the model with the remaining parameters free,
and determining the range for which the $\Delta \chi^2=\chi^2-\chi^2_{\rm
min}<1$ (for 1-$\sigma$ error).

\begin{figure}
 \plotone{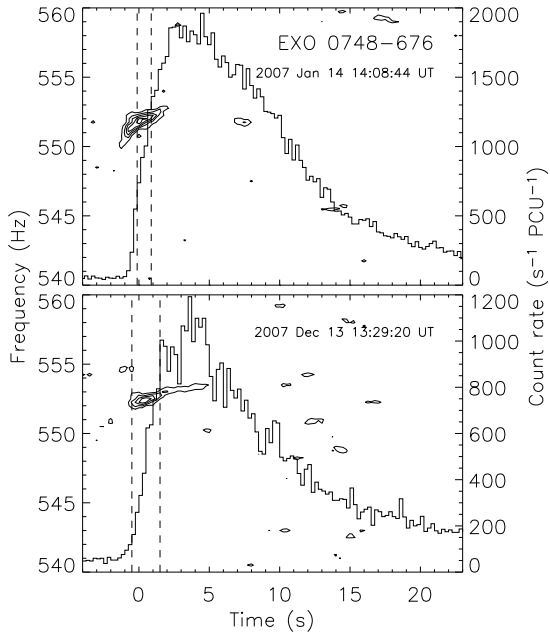}
 \figcaption[]{Dynamic power density spectra computed from oversampled
FFTs covering the two bursts from \src\ in which we detected burst
oscillations. In each panel the contours show the Leahy power as a
function of time and frequency (left-hand $y$-axis). The contour interval
is 10; the overall maximum is slightly lower than in the power spectrum
from the blind search because we oversample by a factor of two. The
histogram shows the burst
(count) profile, binned on 0.25~s (right-hand $y$-axis).
The dashed lines in each panel show the intervals in which the
oscillation was first detected in the blind search.
 \label{pds} }
\end{figure}

To determine the energy dependence of the oscillation we subdivided the
PCA energy band in five intervals adjusted to give approximately the same
source counts in each. We then created background-subtracted lightcurves
for photons within each energy band and folded each lightcurve on the
best-fit pulse frequency determined above. The rms pulse fraction (for a
single sinusoid) increased steadily from $\approx10$\% between 2--4~keV,
to 20\% above 9~keV, consistently in both bursts.  The arrival phase of
the pulse varied weakly below $\approx8$~keV, but appeared to arrive
earlier by 
$0.07\pm0.01$ in phase ($\approx0.1$~ms) above this energy.

We also computed oversampled FFTs covering each burst to measure the
evolution of the oscillation. In both bursts we found evidence of an
increase in the signal frequency, which is characteristic of burst
oscillations (Fig. \ref{pds}). The signal appears initially at a lower
frequency (approximately 551~Hz) in the January 14 burst and
increases in frequency more rapidly.
The overall frequency drift for the two bursts was
no more than 2~Hz (0.4\%).

Following the detection,
we performed a narrowband search for the 
signal in all 154
other bursts detected by \xte\/ from \src, and for which high time
resolution data was available.
As in G08, we computed FFTs of each 1~s interval of
data for the first 16~s of the burst, and searched for signals 
within 5~Hz of the 552~Hz
oscillation frequency.
We considered a signal to be a detection if it had less than a 1\% chance
of occurring due to noise given the 160 trial frequencies searched for
each burst,
and also if it persisted for two adjacent
(independent)
time and frequency bins with a chance probability of $<
(6\times10^{-5})^{1/2}/6 = 1.3\times10^{-3}$, or if it occurred in the
first second of a burst with a chance probability of $<10^{-3}$.
With these criteria we found two additional (albeit weak) detections, in
bursts on 2007 May 19 12:55:55 UT and 2007 Dec  5 12:15:34 UT.
In the May 19 burst oscillations were detected weakly in the peak (defined
as the interval during which the count rate exceeded 90\% of the maximum) and
tail (from the peak onwards), whereas in the December 5 burst the
oscillations were detected in the rise, as with the initial detections.

We also revisited the analysis of \cite{villarreal04}, to measure the
significance of the 45~Hz signal in the full 
sample of 157 bursts observed by \xte\/ from \src\ to date.
We first reproduced the analysis of 
\cite{villarreal04} 
with their original sample of 38 bursts and the same
energy selection and lightcurve rebinning strategy, obtaining similar
results for the 45 Hz peak and its significance.  Next, we applied the
same technique to the larger sample of 157 bursts now available,
but did not detect the signal (using a detection threshold
corresponding to $2\sigma$ single-trial significance).
Finally, noting that the 38 bursts used by \cite{villarreal04}
all occurred at times when the persistent flux 
(measured by the {\it RXTE}/ASM daily average count rates) was atypically
low,
mostly below $1$~count~s$^{-1}$. Therefore, we also searched the subset of
129 bursts from the full sample that also occurred during such low flux
intervals and applied the same search technique. However, the
$45$~Hz signal was again not detected.

\section{Discussion}
\label{sec2}

There are substantial differences in the characteristics of the
two burst oscillations (45~Hz and 552~Hz) now detected in \src.
The 45~Hz signal had a fractional amplitude of $\approx3$\% (rms) and was
detected in the summed power spectrum of 38 bursts,
rebinned by a factor of (typically) 64 or 128 to give a resolution of 1~Hz,
calculated from light curves selecting photons within the energy range
6--60~keV, and covering intervals of typically 64 or 128~s of the burst
decay.
In contrast, the 552~Hz signal has a fractional amplitude of 15\% (rms),
was detected separately in unbinned power spectra calculated from 1- and
2-s light curves extracted over the full
PCA energy range ($\approx2$--60~keV) during the rise of two individual
bursts, separated by 11~months.
Additionally, the dynamic power density spectra give evidence for
frequency evolution while the 552~Hz signal was present, similar to that
observed in other burst oscillation sources.
The key question is, which of these two signals traces the neutron star
spin?

We consider three alternatives. First,
assuming both signals are genuine,
the 45~Hz signal may arise from the neutron star spin,
in which case the 552~Hz
signal may arise from a high-order ($m\approx12$--13) radial mode
(\citealt{cb00}; see also \citealt{heyl04,pb05}). However, the appearance of 
this mode alone is difficult to explain; 
one would expect the excitation of many different
modes, rather than just two with widely-separated orders.
Second, if both signals are real but the 552~Hz signal instead indicates
the neutron star spin, there is no known mechanism 
that can give rise to a 45~Hz signal from the neutron star surface.
As suggested by \cite{balman09}, the 45~Hz oscillation
may instead arise in the boundary layer between the disk and neutron
star.

Third, the lack of a detection of the 45~Hz signal in the larger sample of
bursts detected since 2004 suggests that the 45~Hz signal may have arisen
from statistical fluctuations.
Our analysis shows that including many more bursts in the power spectral
sum 
has the effect only of reducing the detection significance
to well below any conservative detection threshold. The signal power is also
quite sensitive to other parameters of the data selection.
Thus, while it is possible that the 45~Hz oscillation 
is transient and has not been detectable since, it is also
difficult to rule out an origin in statistical noise with any confidence. 

The properties of the 552~Hz oscillation closely resemble those of the
burst oscillations observed in other systems, that are believed to
trace the neutron star spin to within a few Hz.  In contrast, the properties of
the 45~Hz oscillation are quite distinct, especially its low frequency and
the inability to detect it in
individual bursts.
Thus, 
we conclude that the 552~Hz signal almost certainly traces
the neutron star spin in \src; since the signal appears to
increase by up to 2~Hz during the burst rise, we expect that the true spin
frequency may be a few Hz higher.

The energy dependence of the 552~Hz oscillation amplitude in \src\ was
similar to that of other burst oscillation sources \cite[]{muno03c},
although the
overall amplitudes were higher. The suggestion of the hard ($\ga8$~keV)
photons arriving earlier than the soft photons is however contrary to
previous observations. In that respect we note that the previous analysis
focussed principally on oscillations present throughout the tails of
bursts, whereas the oscillations in \src\ were present only in the rise.
It is not known whether the two types of oscillations have
characteristically different variation of pulse arrival time with energy.

The duty cycle of the 552~Hz
oscillation (the number of bursts in which it was detected compared to the
total number observed) is extremely small at $\approx1.2$\%, the smallest
yet for all the burst oscillation sources (e.g. G08).
The unprecedented scarcity of the oscillation raises the question of what
was so unusual about the bursts that exhibited them. Compared to the
entire sample of bursts observed by \xte\/ from \src, the bursts on
January 14 and December 13 had somewhat shorter timescales (calculated as
the ratio of peak flux to fluence) $\tau=12$--13~s, while the
typical range is 15--30~s. Similarly, the rises were of shorter duration
than average, at 2--4~s. Shorter burst timescales suggest a 
smaller proportion of hydrogen in the burst fuel than usual.
Correspondingly, while the fluences were rather typical, the
bursts reached higher than average peak fluxes. However, other bursts were
observed with similar properties which did not exibit oscillations, so
these properties do not uniquely determine their observeability.
The January 14 burst occurred only 11.4~min after the previous event,
and had a fluence only about 77\% lower. Such short recurrence time bursts
are common for \src, and many examples have been detected by \xte\/
(e.g. G08) as well as {\it XMM-Newton}\/ \cite[]{boirin07a}. However, the
second burst in these pairs is typically much fainter than the
first. Interestingly, it is also possible that the December 13 burst was
the second in a closely-spaced pair, as a data gap (due to the 90~min
satellite orbit) prevented observations until approximately 7~min before
the event. Again, timing analysis of other such examples did not reveal
any additional oscillations, however.

We also compared the properties of the persistent emission at the time
when the bursts with oscillations occurred, to other public \xte\/
observations of the source (from G08). The persistent flux in both
observations was $\approx3\times10^{-10}\ \epcs$ (2.5--25~keV),
approximately equal to the 50th percentile value of the flux distribution.
Similarly, the hard and soft spectral colors\footnote{Defined as in G08 as
the ratio of 
the background-subtracted detector counts in the (8.6--18.0)/(5.0--8.6)~keV 
and the (3.6--5.0)/(2.2--3.6)~keV energy bands, respectively} were in the
middle of the observed range, at $\approx0.8$ and $\approx1.6$,
respectively. Thus, we found no evidence for an unusual spectral or
intensity state at the time of the bursts with oscillations.

A 552~Hz spin frequency for \src\ means that spectral features arising
from the neutron star surface will be significantly Doppler broadened, as
well as
reducing the central line depth to only $\la5$\% of the continuum
level \cite[e.g.][]{cbw05}. Narrow lines from such rapidly-rotating
objects can only arise if the system is viewed at extremely low
inclinations; however, the eclipses and dipping activity in \src\
unambiguously indicate a high system inclination. Thus, the narrow
($\approx0.1$~\AA) spectral features reported by \cite{cott02} could not
arise from the neutron star surface. This conclusion is corroborated by
other recent work, including the absence of lines in a deeper {\it
XMM-Newton}\/ spectrum \cite[]{cott08}, as well as difficulties explaining
the inferred Fe column \cite[]{cbw05}. However, 
the narrow spectral features are difficult to identify otherwise
\cite[e.g.][]{kong07}, and a satisfactory explanation remains elusive.

\acknowledgments

We thank Tony Piro, Randy Cooper, Lars Bildsten, Al Levine and Mike Muno
for useful discussions.
This research has made use of data obtained through the High Energy
Astrophysics Science Archive Research Center Online Service, provided by
the NASA/Goddard Space Flight Center.

Facilities: \facility{RXTE}.

\clearpage

\clearpage

\end{document}